\documentclass[12pt]{article}
\usepackage{graphicx}
\usepackage{amsmath}
\usepackage{cite}
\usepackage{amsfonts}
\usepackage{amssymb}
\usepackage{amsmath}
\usepackage{latexsym}
\usepackage{color}

{\catcode `\@=11 \global\let\AddToReset=\@addtoreset}
\AddToReset{equation}{section}

{\catcode `\@=11 \global\let\AddToReset=\@addtoreset}
\AddToReset{figure}{section}

{\catcode `\@=11 \global\let\AddToReset=\@addtoreset}
\AddToReset{table}{section}

\newcommand{\dgamma}{\psi}

\newcommand{\R}{{\mathbf{R}}{}}

\newcounter{mnotecount}[section]

\renewcommand{\themnotecount}{\thesection.\arabic{mnotecount}}

\newcommand{\mnote}[1]
{\protect{\stepcounter{mnotecount}}$^{\mbox{\footnotesize
$
\bullet$\themnotecount}}$ \marginpar{
\raggedright\tiny\em $\!\!\!\!\!\!\,\bullet$\themnotecount: #1}
}

\newcommand{\warn}[1]
{\protect{\stepcounter{mnotecount}}$^{\mbox{\footnotesize
$
\bullet$\themnotecount}}$ \marginpar{
\raggedright\tiny\em $\!\!\!\!\!\!\,\bullet$\themnotecount:
{\bf Warning:} #1} }

\newcommand{\hf}{{\hat f} {}}
\newcommand{\cf}{{\check f}{}}
\newcommand{\alt}{{\hat \alpha} {}}
\newcommand{\alp}{{\check \alpha}{}}
\newcommand{\betat}{{\hat \beta} {}}
\newcommand{\betar}{{\check \beta}{}}

\newtheorem{theorem}{Theorem}[section]

\newtheorem{lemma}[theorem]{Lemma}

\newtheorem{proposition}[theorem]{Proposition}
\newtheorem{Proposition}[theorem]{Proposition}
\newtheorem{remark}[theorem]{Remark}

\newenvironment{proof}[1][Proof]{\textbf{#1.} }{\ \rule{0.5em}{0.5em}}

\begin{document}
%
\title{On free general relativistic initial data on the light cone}

\author{%
    Piotr T. Chru\'{s}ciel \\
    Universit\"at Wien
 \and
    Jacek Jezierski\\
    Uniwersytet Warszawski \\ KMMF, Ho\.za 69, 00-682 Warszawa
}

\maketitle
%

\newcommand{\Rell}{{\stackrel{(\ell)}R}{}}
\newcommand{\psilp}{{\stackrel{(\ell )}\psi}{}}
\newcommand{\dpsilp}{{\stackrel{(\ell )}{\delta\psi}}{}}
\newcommand{\psil}{{\stackrel{(\ell-1)}\psi}{}}
\newcommand{\chil}{{\stackrel{(\ell-1)}\chi}{}}
\newcommand{\dchilp}{{\stackrel{(\ell )}{\delta\chi}}{}}
\newcommand{\chilp}{{\stackrel{(\ell )}\chi}{}}
\newcommand{\gello}{{\stackrel{(1)}g}{}}
\newcommand{\Rello}{{\stackrel{(1)}R}{}}
\newcommand{\Rellz}{{\stackrel{(0)}R}{}}
\newcommand{\gellz}{{\stackrel{(0)}g}{}}
\newcommand{\xellz}{{\stackrel{(0)}x}{}}
\newcommand{\gell}{{\stackrel{(\ell)}g}{}}
\newcommand{\dxell}{\delta {\stackrel{(\ell+3)}x}{}}
\newcommand{\xell}{{\stackrel{(\ell)}x}{}}
\newcommand{\dgell}{{\stackrel{(\ell+2)}{\delta g}}{}}
\newcommand{\dgellm}{{\stackrel{(\ell+1)}{\delta g}}{}}
\newcommand{\Gammal}{{\stackrel{(\ell)}\Gamma}{}}
\newcommand{\Gammalp}{{\stackrel{(\ell+1)}\Gamma}{}}
\newcommand{\Rellp}{{\stackrel{(\ell+1)}R}{}}
\newcommand{\gellp}{{\stackrel{(\ell+1)}g}{}}
\newcommand{\xellp}{{\stackrel{(\ell+1)}x}{}}
\newcommand{\gellm}{{\stackrel{(\ell-1)}g}{}}
\newcommand{\gone}{{\stackrel{(1)}g}{}}

\newcommand{\usigiy}{\stackrel{(i+2)}\underline{\sigma}_{\!\!\! jk}}
\newcommand{\usigi}{\stackrel{(i+2)}{\underline{\sigma}}_{\!\!\! AB}}
\newcommand{\divsigi}{{\mathcal D}^j \stackrel{(i+2)}{\underline{\sigma}}_{\!\!\! j k}}
\newcommand{\fif}{\stackrel{(i+2)}f}
\newcommand{\gi}{\stackrel{(i+2)}g}
\newcommand{\fm}{\stackrel{(m)}f}
\newcommand{\gm}{\stackrel{(m)}g}
\newcommand{\fmpt}{\stackrel{(m+2)}f}
\newcommand{\gmpt}{\stackrel{(m+2)}g}
\newcommand{\usigmy}{\underline{\stackrel{(m)}\sigma}_{\!\!ij}}
\newcommand{\usigm}{\underline{\stackrel{(m)}\sigma}_{\!\!\! AB}}
\newcommand{\divsigm}{{\mathcal D}^j \underline{\stackrel{(m)}\sigma}_{\!\!\! ij}}
\newcommand{\mcD}{{\mathcal D}}

\newcommand{\myw}{x}
\newcommand{\regular}{w}

\newcommand{\zD}{{\mathring \nabla}}

\begin{abstract}
 We provide a simple explicit parameterization of free general
 relativistic data on the light cone.
\end{abstract}

\tableofcontents

\section{Introduction}

In a series of recent papers~\cite{CCM2,CCM3,CCM4,CMG},
solutions of the vacuum Einstein equations defined to the
future of a light cone, say $C_O$, issued from a point $O$,
have been characterized in terms of data on a light cone. Part
of those data is provided by a symmetric degenerate tensor on
$C_O$, and the approach there requires this degenerate tensor
to be induced on $C_O$ by \emph{some} smooth Lorentzian metric
$C=C_{\mu\nu}dx^\mu dx^\nu$. The question then arises, how to
usefully describe the induced tensors  having this property.
Now, tensor fields on $(0,R)\times S^2$ with vanishing
$r$--components, where $r$ parameterizes $(0,R)$, can always be
written in the form (see, e.g., \cite[Appendix~E]{JJpeeling})
\begin{eqnarray}
 \label{18IX0.1}
 \hspace*{-0.3cm}
 r^2 \big[(1+ \gamma)\mathring s_{AB}+ 2\alpha_{||AB}
  - \mathring s_{AB} \mathring s^{CD} \alpha _{||CD}
  + \mathring \epsilon_A{}^C  \beta_{||CB}
  + \mathring \epsilon_B{}^C  \beta_{||CA}\big] dx^A dx^B
  \,   ,
\end{eqnarray}
where  $\mathring s\equiv \mathring s_{AB} dx^A dx^B$ is the
round unit metric on $S^{2}$, and $||$ denotes covariant
differentiation on $(S^{2},\mathring s)$. Further, $\mathring
s^{AB}$ is the inverse metric to $\mathring s_{AB}$, $\mathring
\epsilon^A{}_B:= \mathring s^{AC} \mathring \epsilon_{CB}$, and
$\mathring \epsilon_{AB}$ is the alternating tensor on
$(S^2,\mathring s)$. This shifts the extendibility question to
that of the properties of the functions $\alpha$, $\beta$ and
$\gamma$. The aim of this note is to prove the following (see
Section~\ref{ss14IX0.1} for terminology and
Section~\ref{ss15IX0.1} for the proof):

\begin{theorem}
 \label{T18IX0.1}
A tensor field on $(0,R)\times S^2$  of the form
\eqref{18IX0.1} is the restriction of a smooth metric in normal
coordinates to its light cone if and only if the functions
$\alpha$, $\beta$ and $\gamma$ are $C_O$--smooth, except
possibly for the $\ell=0$ and $\ell=1$ spherical harmonics of
$\alpha$ and $\beta$ which give zero contribution to
\eqref{18IX0.1}.
\end{theorem}

Consider the vacuum  general relativistic characteristic
constraint equation in the affinely parameterized gauge (see,
e.g., \cite{CCM2}):
\begin{equation}
\partial_{1}\tau +\frac{\tau^{2}}{n-1}+|\sigma|^{2} =0
\;,
 \label{9IV10.1x}
\end{equation}
where $\tau$ is the divergence of $C_O$ and $\sigma$ its shear.
Given $\alpha$ and $\beta$, Equation~\eqref{9IV10.1x} can be
viewed as a non-linear ODE for $\gamma$, and thus the functions
$\alpha$ and $\beta$ can be thought of as representing
unconstrained degrees of freedom of the gravitational field.
%

Theorem~\ref{T18IX0.1} invokes normal coordinates for the
metric $C$, and its proof requires a useful description of the
components of a metric tensor in normal coordinates. This is
provided by the following result, proved in
Section~\ref{ss14IX0.1}, which has some interest of its own:

\begin{theorem}
 \label{T18IX0.2}
 The coordinates $w^\mu$ are normal for a metric $C_{\mu\nu}$  if and
 only if there exists a tensor field
 $\Omega_{\alpha\beta\gamma\delta} $ satisfying
\begin{equation}
 \label{10IX0.3}
 \Omega_{\alpha\beta\gamma\delta} = \Omega_{\gamma\delta \alpha\beta} = - \Omega_{\beta\alpha\gamma\delta}
\end{equation}
 such that
\begin{equation}
 \label{18IX0.4}
 \underline{g_{\alpha\gamma}} =
  \underline{ \eta_{\alpha\gamma}} +\underline{
 \Omega_{\alpha\beta\gamma\delta}} w^\beta w^\delta
  \;,
\end{equation}
where underlined tensor components denote coordinate components
in the coordinate system $w^\mu$, and where $\eta$ is the
Minkowski metric.
\end{theorem}

\begin{remark}
 \label{R10IX0.1}
{\rm While we are mainly interested in Lorentzian metrics, we
note that Theorem~\ref{T18IX0.2} has a direct counterpart in
all signatures.
}
\end{remark}

The main issue of our work is the understanding of the
behaviour of the objects at hand near the vertex of the cone.
Many of the considerations below are valid only within the
domain of definition of normal coordinates centered at the
vertex of the light cone, which is sufficient for the purpose.

\section{Tensors and the light cone}

Consider a smooth metric $C$ in normal coordinates
$\regular^\mu$. As already pointed out, we write
$\underline{C_{\beta \gamma }}$ for the coordinate components
of the metric tensor in this coordinate system. We reserve the
notation  $C_{\mu\nu}$ for the components of $C$ in the
coordinate system $(x^0\equiv u,x^1 \equiv r,x^A)$, defined as
\begin{equation}
\regular^{0}=x^{1}-x^{0},
\qquad
\regular^{i}=x^1\Theta^{i}(x^{A})
\qquad \text{with} \quad
\sum_{i=1}^{n}\left[\Theta^{i}(x^{A})\right]^{2}=1
 \;.
 \label{19XI.3}
\end{equation}
Thus
$$
 \partial_u = - \partial_{w^0}\;,
 \quad
 \partial_r = \partial_{w^0} + \frac{w^i}r \partial_{w^i}
 \;,
$$
and
$$
 \eta = -(dx^0)^2 + 2 dx^0 dx^1 + r^2 \mathring s_{AB} dx^A dx^B\;,
 \quad
 \eta^\sharp = \partial_r^2 + 2 \partial_ u \partial_ r + r^{-2} \mathring s^{AB}\partial_ A \partial_ B
 \;.
$$

The explicit form of the transformation formulae for a
symmetric tensor $T_{\mu\nu}$ reads
\begin{equation}
 \label{22XI.21}
T_{00}\equiv\underline{T_{00}},\quad
T_{01}\equiv-\underline{T_{00}} -\underline{T_{0i}}\Theta^{i},\quad
T_{0A}\equiv-\underline{T_{0i}} r\frac{\partial\Theta^{i}}{\partial x^{A}}
 \;,
\end{equation}
\begin{equation}
 \label{22XI.22}
T_{11}\equiv\underline{T_{00}}+2\underline{T_{0i}}\Theta^{i}
           +\underline{T_{ij}}\Theta^{i}\Theta^{j},\quad
T_{1A}\equiv\underline{T_{0i}} r\frac{\partial\Theta^{i}}{\partial x^{A}}
      +\underline{T_{ji}}r\Theta^{j}
          \frac{\partial\Theta^{i}}{\partial x^{A}}\;,
\end{equation}
\begin{equation}
 \label{22XI.23}
T_{AB}\equiv\underline{T_{ij}}r^{2}\frac{\partial\Theta^{i}}{\partial x^{A}%
}\frac{\partial\Theta^{j}}{\partial x^{B}} \;.
\end{equation}
Conversely, $\underline{T_{\lambda\mu}}=\frac{\partial x^{\alpha}%
}{\partial \regular^{\lambda}}\frac{\partial x^{\beta}}{\partial \regular^{\mu}}%
T_{\alpha\beta}$ gives
\begin{equation}
 \label{20XII.1}
\underline{T_{00}}\equiv T_{00},\quad
\underline{T_{0i}}\equiv
-(T_{00}+T_{01})\Theta^{i}-T_{0A}\frac{\partial x^{A}}{\partial \regular^{i}}
\;,  %
\end{equation}
\begin{equation}
 \label{20XII.2}
\underline{T_{ij}}=(T_{00}+2T_{01}+T_{11})\Theta^{i}\Theta^{j}
+(T_{0A}+T_{1A})
\left(\Theta^{i}\frac{\partial x^{A}}{\partial \regular^{j}}+
\Theta^{j}\frac{\partial
x^{A}}{\partial \regular^{i}}\right)
+T_{AB}\frac{\partial x^{A}}{\partial \regular^{i}}\frac{\partial x^{B}}{\partial \regular^{j}}
 \;.
\end{equation}

An overline over a function $f$ denotes restriction of the
function to the light cone $C_O=\{w^0 = |\vec w|\}$: if we
parameterize the cone by $\vec w \equiv (w^i)$, we have
$$
\overline {f} (\vec w) := f(w^0 =|\vec w |, \vec w)
 \;,
$$
where $|\vec w|^2 :=  \sum_{i} (w^i)^2$.

Note that the domain of definition of normal coordinates for a
general metric is rarely global, and that our considerations
apply only within this domain.

Since $C_O$ can be coordinatised by $\vec w$, functions on
$C_O$ can be identified with functions of $\vec w$. A function
$\varphi$ on $C_O$ will be said to belong to $C^k(C_O)$ if
$\varphi$ can be written as $\hat \varphi +r \check \varphi$,
where $\hat \varphi$ and $\check \varphi$ are $C^k$ functions
of $\vec w$. A function on $C_O$ will be called
\emph{$C_O$--smooth} if it can be written as  $\hat \varphi +r
\check \varphi$, where $\hat \varphi$ and $\check \varphi$ are
smooth functions of $\vec w$. A similar definition is used for
real-analytic functions. It is not too difficult to show that a
function $\varphi$ is $C_O$--smooth if and only if there exists
a smooth function $\varphi$ on space-time such that
$\varphi=\overline f$. In other words:

\begin{Proposition}
 \label{P18IX0.1} A function $\varphi$ defined on
$$
 C_O:= \bigg\{w^\mu \in {\mathbf{R}}^{n+1}:\ w^0 = \sqrt{\sum_ i (w^i)^2}\bigg\}
$$
can be extended to a $C^k$, respectively smooth, respectively
analytic, function on ${\mathbf{R}}^{n+1}$ if and only if
$\varphi$ is $C^k(C_O)$, respectively $C_O$--smooth,
respectively $C_O$--analytic.
\end{Proposition}

The proof  of Proposition~\ref{P18IX0.1}  for real-analytic
functions can be found in~\cite{CCM4}; the remaining cases are
covered in Appendix~\ref{A22IX0.1}.

\subsection{Normal coordinates}
 \label{ss14IX0.1}

Recall that (local) coordinates $w^\mu$ are normal for the
metric $C$ if and only if it holds that~\cite{Thomas}
\begin{equation}
 \label{19XI.1}
 \underline{ C_{\mu\nu}}\regular^\mu = \underline{\eta_{\mu\nu}} \regular^\mu
 \;.
\end{equation}
For completeness, and because of restricted accessibility
of~\cite{Thomas}, we give a proof of this in
Appendix~\ref{A19IX0.1}.

It follows from \eqref{22XI.21} and \eqref{19XI.1} that
\begin{eqnarray}
 \label{6IX0.1}
 \overline{C_{11}} &=&
  \frac 1 {r^2} \overline{\underline{C_{\mu\nu}w^\mu w^\nu}}
 = \frac 1 {r^2} \overline{\underline{\eta_{\mu\nu}}w^\mu w^\nu} = 0\;,
\\
 \label{6IX0.2}
 \overline{C_{01}}
  &= &
  -\frac 1 {r} \overline{\underline{C_{0\nu}} w^\nu}
 = -\frac 1 {r} \overline{\underline{\eta_{0\nu}} w^\nu}
  = 1\;,
\\
 \nonumber
 \overline{C_{1A}} &= &
           \overline{ \underline{C_{i\mu}}w^\mu\frac{\partial\Theta^{i}}{\partial x^{A}}}
\\
 \label{6IX0.3}
 &= & \overline{ \underline{\eta_{i\mu}}w^\mu\frac{\partial\Theta^{i}}{\partial x^{A}}}
 =  r\sum_i  { \Theta^i\frac{\partial\Theta^{i}}{\partial x^{A}}}
 =  \frac 12 \sum _i r {\frac{\partial(\Theta^{i} \Theta^i)}{\partial x^{A}}}
 =0
\end{eqnarray}
(note that the only information, that does not immediately
follow from the fact that $
 C_O $
is  the future light cone for the metric $C$, is provided
by\eqref{6IX0.2}; the remaining equations can serve as
consistency checks).

We set %
$$
 h_{\mu\nu}:= C_{\mu\nu} -\eta_{\mu\nu}
 \;,
$$
and we will lower and raise all indices with the metric $\eta$.
Hence the coordinates $w^\alpha$ are normal for
$C=C_{\mu\nu}dw^\mu dw^\nu$ if and only if
\begin{equation}
 \label{3IX0.1}
 \underline{h_{\mu\nu}}w^\mu = 0
 \;.
\end{equation}
Note that, from \eqref{6IX0.1}-\eqref{6IX0.3},
\begin{equation}
 \label{6IX0.4}
 \overline{h_{1\mu}}   = 0
  \quad
   \Longleftrightarrow
    \quad
 \overline{h^{0\mu}} := \eta^{0\alpha} \eta^{\mu \beta} h_{\alpha \beta}  = 0
 \;.
\end{equation}

The question arises, how to describe exhaustively, and in a
useful way, the set of tensors satisfying \eqref{3IX0.1}. One
obvious way of doing this is to use a projection operator:
indeed, for any smooth symmetric tensor $\phi_{\mu\nu}$, the
tensor field
$$
 P_{\alpha}{}^\mu
 P_{\beta}{}^\nu (\eta_{\rho\sigma} w^\rho w^\sigma)^2\phi_{\mu\nu}\;, \ \mbox{where} \
 P_{\alpha}{}^\beta  = \delta_ \alpha^\beta - \frac{\eta_{\alpha\mu}w^\mu w^\beta}{\eta_{\rho\sigma} w^\rho w^\sigma}
$$
is a smooth tensor field satisfying \eqref{3IX0.1}. This leads
to a restricted class of tensors because of the multiplicative
factor $(\eta_{\rho\sigma} w^\rho w^\sigma)^2$ above (in
particular the resulting tensor induced on the light cone has
vanishing $AB$ components), and it is not clear how to
guarantee smoothness of the final result without the
multiplicative factor. Variations on the above using a space
projector $\delta^i_j - r^{-2} x^i x^j$ lead to similar
difficulties.

Note, however, that solutions of \eqref{3IX0.1} can be
constructed as follows: let $\Omega_{\alpha\beta\gamma\delta}$
be any smooth tensor field satisfying \eqref{10IX0.3}. Then the
tensor field
\begin{equation}
 \label{10IX0.1}
 \underline{h_{\alpha\gamma}} =
  \underline{
 \Omega_{\alpha\beta\gamma\delta}} w^\beta w^\delta
\end{equation}
is symmetric, and satisfies \eqref{3IX0.1}.
Theorem~\ref{T18IX0.2} follows now immediately from:

\begin{proposition}
 \label{P10IX0.1}
 A tensor field $h_{\mu\nu}$ satisfies \eqref{3IX0.1} if and
 only if there exists a tensor field
 $\Omega_{\alpha\beta\gamma\delta} $ satisfying \eqref{10IX0.3}
 such that \eqref{10IX0.1} holds.
\end{proposition}

\noindent\proof
We work in a given smooth coordinate system $ x^\mu$. The
sufficiency has already been established. To show necessity
recall, first, that any smooth tensor field satisfying
\begin{equation}
 \label{10IX0.2}
 A_ \mu x^\mu =0
\end{equation}
can be represented as
$$
 A_{\mu} = \Omega_{\mu \nu} x^\nu\;, \ \mbox{with} \ \Omega_{\mu\nu}=-\Omega_{\nu\mu}
  \;.
$$
To see this, note first that differentiation of \eqref{10IX0.2}
shows that $A_\mu(0)=0$; then
\begin{eqnarray*}
 A_ \mu (x^\sigma) &= &  \int_0^1 \frac{d}{ds}\left[
 s A_ \mu(s x^\sigma) \right] ds =
 \int_0^1  \left[
  A_{\mu}( s x^\sigma) + s x^\nu \partial_ \nu A_ \mu(s x^\sigma)  \right] ds
\\
 & = &
x^\nu \underbrace{ \int_0^1   s(\partial_ \nu A_ \mu - \partial_ \mu A_ \nu) (s x^\sigma)   ds}_{=: \Omega_{\mu \nu}}
 \;,
\end{eqnarray*}
where we have used
$$
 \partial_\mu( x^\nu A_\nu)=0 \quad \Longrightarrow \quad A_\mu (s x^\sigma)= -s x^\nu \partial_ \mu A_\nu (s x^\sigma)
 \;.
$$
Applying this to $h_{\mu\nu}$ at fixed $\nu$ we find that there
exists a field $\Omega_{\alpha\beta\nu}$, anti-symmetric in
$\alpha$ and $\beta$, such that
$$
 h_{\mu\nu}(x^\rho)  = \Omega_{\mu \alpha \nu}(x^\rho)  x^\alpha
 \;.
$$
Applying the construction again to the last equation at fixed
$\mu$ and $\alpha$ we conclude that
$$
 \Omega_{\mu \alpha \nu}(x^\rho)  =
 \Omega_{\mu \alpha \nu\beta}(x^\rho) x^\beta
$$
for some  field $\Omega_{\alpha\beta\gamma\delta}$,
anti-symmetric in $\gamma$ and $\delta$. This is of the desired
form, but the pair-interchange symmetry is not completely
clear. However, the above prescription gives
\begin{eqnarray}
\lefteqn{
 \Omega_{\mu\sigma\nu\lambda}(x^\rho) =
  }
  &&
   \nonumber
\\
  &&
    \int_0^1 s^2 \int _0^1
  t \left(
   \partial_\lambda \partial_\sigma h_{\mu\nu} -
   \partial_\lambda \partial_\mu h_{\sigma\nu} +
   \partial_\nu \partial_\mu h_{\sigma\lambda} -
   \partial_\nu \partial_\sigma h_{\mu\lambda}
   \right)(stx^\rho)\, dt\, ds
   \;,
 \nonumber
\\
 &&
    \label{15IX0.3}
\end{eqnarray}
which makes manifest all the symmetries claimed. This equation
defines the components of the tensor field
$\Omega_{\mu\sigma\nu\lambda}(x^\rho)$ in the coordinate system
$x^\mu$.
\hfill$\Box$

One should bear in mind that $\Omega_{\alpha\beta\gamma\delta}$
is not uniquely defined by \eqref{10IX0.1}. However,
\eqref{15IX0.3} can be used as a canonical choice, if
needed.

\bigskip

It would be of interest to provide an answer to the corresponding
question for tensor fields satisfying \eqref{3IX0.1} on the
light cone only:
\begin{equation}
 \label{3IX0.1x}
 \overline{
 \underline{h_{\mu\nu}}w^\mu} = 0
 \;.
\end{equation}
We return to this question in Section~\ref{s22IX0.1}, where
some partial results are given, but we have not attempted an
exhaustive study. In any case, on the light cone \eqref{3IX0.1}
gives the following:
\begin{eqnarray}
 \label{10IX0.4}
 \overline{\underline{h_{00}}}
  & = &
 \overline{\underline{\Omega_{0i0j}}} w^i w^j
 \;,
\\
 \label{10IX0.5}
 \overline{\underline{h_{0i}}}
  & = & \big(-
 \overline{\underline{\Omega_{0j0i}}} r
  +
 \overline{\underline{\Omega_{0jik}}} w^k \big) w^j
 \;,
\\
 \label{10IX0.6}
 \overline{\underline{h_{ij}}}
  & = &
 \overline{\underline{\Omega_{i0j0}}}r^2
  +
   \big[
   -
 \overline{\underline{\Omega_{0ijk}}}r
  -
 \overline{\underline{\Omega_{0jik}}}r
  +
 \overline{\underline{\Omega_{ikj\ell}}} w^\ell
  \big]w^k
 \;.
\end{eqnarray}
In coordinates adapted to the light cone
\eqref{10IX0.4}-\eqref{10IX0.6} translate to
\begin{eqnarray}
 \label{10IX0.7}
 \overline{{h_{00}}}
  & = &
 \overline{\underline{\Omega_{0i0j}}} w^i w^j
 \;,
\\
 \label{10IX0.8}
 \overline{{h_{\mu 1}}}
  & = &
   0
   \;,
\\
 \label{10IX0.9}
 \overline{{h_{0A}}}
  & = & r\big(
 \overline{\underline{\Omega_{0j0i}}} r
  -
 \overline{\underline{\Omega_{0jik}}} w^k \big) w^j\frac{\partial \Theta^i}{\partial x^A}
 \;,
\\
 \overline{{h_{AB}}}
 & = &
 r^2
  \bigg(
 \overline{\underline{\Omega_{i0j0}}}r^2
 \nonumber
\\
 \label{10IX0.10}
  &&
   +
   \big(
   -
 \overline{\underline{\Omega_{0ijk}}}r
  -
 \overline{\underline{\Omega_{0jik}}}r
  +
 \overline{\underline{\Omega_{ikj\ell}}} w^\ell
  \big)w^k
   \bigg)
   \frac{\partial \Theta^i}{\partial x^A}
   \frac{\partial \Theta^j}{\partial x^B}
 \;.
\end{eqnarray}
In particular $\overline{{h_{0A}}}$ factors out through $r$ and
is $O(r^3)$, while $\overline{{h_{AB}}}$ factors out through
$r^2$ and is $O(r^4)$.

For further use we note
\begin{eqnarray}
 \nonumber
  \lefteqn{
   \overline{{h_{AB}}}
   \frac{\partial x^A}{\partial w^p}
   \frac{\partial x^B}{\partial w^q}
   }
   &&
\\
  & = &
  {
 \overline{\underline{\Omega_{i0j0}}}(r  \delta^ i  _p - w^i
 \Theta^p) (r  \delta ^j _ q - w^j \Theta^q)
  }
 \nonumber
\\
  &&
   +
   \big[
   -
 \overline{\underline{\Omega_{0iqk}}}(r \delta^ i  _p - w^i
 \Theta^p)
  -
 \overline{\underline{\Omega_{0jpk}}}(r \delta ^j _ q - w^j \Theta^q)
  +
 \overline{\underline{\Omega_{pkq\ell}}} w^\ell
  \big]w^k
 \nonumber
\\
 &= &
 \overline{\underline{\Omega_{i0j0}}}  w^i
  w^j \Theta^p  \Theta^q
   +
   w^k  w^i\big(
 \overline{\underline{\Omega_{0iqk}}}
 \Theta^p
 +
 \overline{\underline{\Omega_{0ipk}}}    \Theta^q
  \big) \label{12IX0.1}
\\
  &&
  +r^2\overline{\underline{\Omega_{p0q0}}} -
 \overline{\underline{\Omega_{i0q0}}}  w^i
 w^p -
 \overline{\underline{\Omega_{p0i0}}}  w^i    w^q
   -
   \big(
 r\overline{\underline{\Omega_{0pqk}}}
  +
 r\overline{\underline{\Omega_{0qpk}}}
 -
 \overline{\underline{\Omega_{pkq\ell}}} w^\ell
  \big)w^k
 \;. \nonumber
\end{eqnarray}
This equation has been derived under the assumption that the
coordinates $w^\mu$ are normal; however, $\overline{h_{AB}}dx^A
dx^B$ is intrinsic to the light cone, and hence this equation
provides the most general form of a tensor field
$\overline{h_{AB}}dx^A dx^B$ arising from some smooth metric
$C_{\mu\nu}$ in coordinates which coincide with the normal ones
on the light cone $C_O$.

Note that given the specific structure of the terms containing
$\Theta^i$ above, it is clear how to extract
$\overline{\underline{\Omega_{0i 0j}}}$ and
$\overline{\underline{\Omega_{0ijk}}}$ from $h_{AB}$.

We shall say that a tensor field $h$ is $C_O$--smooth if there
exists a coordinate system $w^\mu$ in which the components of
$h$ are $C_O$--smooth. We conclude that (keeping in mind the
local character of normal coordinates):

\begin{Proposition}
 \label{P14IX0.1}
A tensor field ${\varphi_{AB}}dx^A dx^B$  on $C_O$ arises from
the restriction to the light cone of a metric in normal
coordinates if and only if there exist  $C_O$--smooth tensor
fields $A_{ij}$, symmetric in its indices, $A_{ijk}$,
anti-symmetric in the last two indices, and $A_{ijkl}$,
satisfying $A_{ijkl}=A_{klij}=-A_{jikl}$, such that
\begin{eqnarray}
 \nonumber
  \lefteqn{\left(
   {\varphi_{AB}} - r^2\mathring s_{AB}\right)
   \frac{\partial x^A}{\partial w^p}
   \frac{\partial x^B}{\partial w^q}
   }
   &&
\\
  & = &
 \underline{A_{ij}}  w^i
  w^j \Theta^p  \Theta^q
   +
   w^k\big(
 \underline{{A_{iqk}}}  w^i
 \Theta^p
 +
 \underline{{A_{jpk}}}  w^j \Theta^q
  \big)
 \nonumber
\\
  &&
  +r^2\underline{{A_{pq}}} -
 \underline{{A_{iq}}}  w^i
 w^p -
 \underline{{A_{pi}}}  w^i    w^q
   -
   \big(
 r\underline{{A_{pqk}}}
  +
 r\underline{{A_{qpk}}}
 -
 \underline{{A_{pkq\ell}}} w^\ell
  \big)w^k
 \, .
 \phantom{xx}
 \label{14IX0.1}
\end{eqnarray}
\end{Proposition}

\noindent\proof
The necessity is clear from \eqref{12IX0.1}. To show
sufficiency, suppose that a tensor field satisfying
\eqref{14IX0.1} is given. Let $\Omega_{\mu\nu\rho\sigma}$ be
any smooth tensor field satisfying
$\Omega_{\mu\nu\rho\sigma}=-\Omega_{\nu\mu\rho\sigma}=\Omega_{\rho\sigma\mu\nu}$
such that
$$
 \overline{\underline{\Omega_{0i0j}}} = \underline{A_{ij}}
  \;,
  \quad
 \overline{\underline{\Omega_{0ijk    }}} = \underline{A_{  ijk}}
  \;,
  \quad
 \overline{\underline{\Omega_{ijkl}}} = \underline{A_{ijkl}}
  \, ;
$$
existence of $\Omega_{\alpha\beta\gamma\delta}$ follows from
Proposition~\ref{P18IX0.1}. Then ${\varphi_{AB}}$ is the
restriction to the light cone of the smooth tensor field
$\underline{\eta_{\mu\nu}} +
\underline{\Omega_{\mu\rho\nu\sigma}} w^\rho w^\sigma$ for
which the coordinates $w^\mu$ are normal.
\hfill$\Box$

\bigskip

Recall~\cite{CCM2} (compare~\cite{RendallCIVP,RendallCIVP2})
that solutions of the Cauchy problem for the vacuum Einstein
equations with initial data on an affinely-parameterized light
cone are uniquely determined by the conformal class of
$\overline{C_{AB}}dx^A dx^B$. The remaining components of
$C_{\mu\nu}$ are thus irrelevant for that purpose, and for the
sake of computations it is convenient to choose them as simple
as possible. It is therefore of interest to enquire whether any
$\overline{C_{AB}}$ can be realized by a smooth metric
satisfying
\begin{equation}
 \label{20XI.3}
 \overline{\underline{C_{00}}}= -1\;,\quad
 \overline{\underline{C_{0i}}}= 0 \;, \quad
 \overline{\underline{C_{ij}}}w^j = w^i
 \;.
\end{equation}
Our equations above show that this is only possible for
$C_{AB}$'s which, in coordinates which coincide with the normal
ones on $C_O$, are of the form
\begin{eqnarray}
 \nonumber
 \overline{{h_{AB}}}
   \frac{\partial x^A}{\partial w^p}
   \frac{\partial x^B}{\partial w^q}
 & = &
 \overline{\underline{\Omega_{pkq\ell}}} w^\ell
   w^k
 \;.
 \label{12IX0.2}
\end{eqnarray}
Equivalently, all the functions $ \nonumber
 \overline{{h_{AB}}}
   \frac{\partial x^A}{\partial w^p}
   \frac{\partial x^B}{\partial w^q}$ are $C_O$--smooth.

 \bigskip

We finish this section by the following curious observation,
which shows that normal coordinates can be induced from
one-dimension-up:

\begin{proposition}
 \label{P12IX0.1}
The coordinates $w^i|_{w^0 =0}$ are normal for the metric
$$
 g_{ij}|_{w^0=0} dw^i dw ^j
 \;.
$$
\end{proposition}

\noindent \proof
From $h_{\mu\nu}w^\mu =0$ one finds $h_{ij}|_{w^0=0}
w^i=0$, and the result follows from the Riemannian counterpart
of the equivalence \eqref{3IX0.1}.
\hfill$\Box$

\subsection{Scalar potentials for the metric in dimension $3+1$}
 \label{ss15IX0.1}

So far we have been using general space dimension $n$. For
$n=3$, using a standard decomposition (cf., e.g.,
\cite{JJpeeling}) of symmetric tensors on $S^{n-1}=S^2$ we can
write
\begin{eqnarray}
 \label{13IX0.4}
 \overline{C_{AB}} = r^2 \big[(1+ \gamma)\mathring s_{AB}+ 2\alpha_{||AB}
  - \mathring s_{AB} \mathring s^{CD} \alpha _{||CD}
  + \mathring \epsilon_A{}^C  \beta_{||CB}
  + \mathring \epsilon_B{}^C  \beta_{||CA}\big]
     ,
   \phantom{x}
\end{eqnarray}
We wish to find necessary and sufficient conditions on the
functions $\alpha$, $\beta$ and $\gamma$ so that
$\overline{C_{AB}}$ arises from a smooth metric on space-time.

For reasons that will become apparent shortly, we want to
calculate
$$
 \eta^{\alpha \mu} \eta^{\beta \nu}  \zD_\alpha \zD_\beta C_{\mu\nu}
  \ \mbox{ and } \
 \eta^{\sigma \rho} T_\alpha w_\beta
 \epsilon^{\alpha \beta\gamma\delta} \zD_\rho\zD_\gamma C_{\delta\sigma}
 \;,
$$
where $\zD$ is the covariant derivative of the metric $\eta$,
while
$$
 \underline{T_\alpha} := \underline{\eta_{\alpha 0}} =
 -\delta ^0 _\alpha\;, \quad  \underline{w_ \alpha} =
 \underline{\eta_{\alpha\beta} w^\beta}
 \;.
$$

The calculation of $\eta^{\alpha \mu} \eta^{\beta \nu}
\zD_\alpha \zD_\beta C_{\mu\nu}$ can, and will, be done without
assuming $n=3$; we will use the symbol $\mathring s$ to denote
the unit round metric on $S^{n-1}$. Writing $(x^a) = (x^0,
x^1)$, from
$$
 \mathring \Gamma^A_{rB} = \frac 1 r \delta^A _B\;, \quad \mathring \Gamma^u _{AB} = - \frac 1 r \eta_{AB}=
 \mathring \Gamma^r _{AB}
 \;,
$$
we find
\begin{eqnarray*}
 \zD_\mu h^{\mu\nu} & := & \eta^{\mu\sigma} \eta^{\nu \beta}
 \zD_\sigma h_{\alpha \beta}
\\
 & = &
  \partial_ A h^{A\nu} + \partial _ ah^{a\nu} + 2 h^{rB} \mathring \Gamma ^\nu _{Br}
  + \frac {n-1} r h^{\nu r} + h^{\nu A} \mathring\Gamma^B _{AB}
   + h^{AB}\mathring \Gamma^\nu _{AB}
 \;.
\end{eqnarray*}
Hence
\begin{eqnarray}
 \zD_\mu h^{\mu b}
 & = &
   h^{A b}{}_{||A} + \partial _ ah^{a b}
  + \frac {n-1} r h^{b r}
  - \frac 1r h_{AB} \eta ^{AB}
 \;,
  \label{6IX0.9}
\end{eqnarray}
where $||$ denotes covariant differentiation on
$(S^{n-1},\mathring s)$. Further, using $\zD_\mu X^\mu = |\det
\eta|^{-1/2}\partial_ \mu (|\det \eta|^{1/2}X^\mu)$,
\begin{eqnarray}
 \nonumber
 \eta^{\alpha \mu} \eta^{\beta \nu}  \zD_\alpha \zD_\beta h_{\mu\nu}
 \!\!\! &= & \!\!\!
 h^{AB}{_{||AB}}+h^{ab}{_{,ab}}+2\partial_a h^{aA}{_{||A}}+
 \frac{n+3} r h^{rA}{_{||A}}+
  \frac{n+1}{r} \partial_ a h^{ra}
\\
 &&
  -\frac1r (\partial_u H + \partial_r H)-
 \frac1{r^2} H +\frac{n-1}{r^2}h^{rr}
  \, ,
   \label{4IX0.3}
\end{eqnarray}
and
$$
 H:=\eta^{AB}h_{AB}\;, \ \mbox{ hence } \ \overline{H} =\overline{ \eta^{\mu\nu} h_{\mu\nu}}
  \;.
$$
%
%

To analyze the right-hand side of \eqref{4IX0.3} the following
formulae are useful:
\begin{eqnarray}
 h^{rr}
   &=  &
    h_{uu} +2 h_{ur} +h_{rr} =  \underline{h_{ij}} \Theta^i \Theta^j
  \;,
\\
 h^{ur}
   &=  &
    h_{ur} + h_{rr} = \underline{h_{0i}}\Theta^{i}
           +\underline{h_{ij}}\Theta^{i}\Theta^{j}
  \;,
\\
 h^{uA}
   &=  &
    h_{rA} = \underline{h_{0i}} r\frac{\partial\Theta^{i}}{\partial x^{A}}
      +\underline{h_{ji}} w^{j}
          \frac{\partial\Theta^{i}}{\partial x^{A}}
  \;,
\\
 h^{uu}
   &=  &
    h_{rr} = \underline{h_{00}}+2\underline{h_{0i}}\Theta^{i}
           +\underline{h_{ij}}\Theta^{i}\Theta^{j}
            \;,
\\
 H & \equiv & \eta^{AB}h_{AB} =  \eta^{\mu\nu}h_{\mu\nu} +\underline{h_{00}}
           -\underline{h_{ij}}\Theta^{i}\Theta^{j}
 \;.
\end{eqnarray}
Functions of the form $r^{-2}(\mu + r \nu)$, where $\mu$ and
$\nu$ are restrictions to the light cone of smooth functions on
space-time, will be called \emph{mildly singular}.  In what
follows one should keep in mind that any function $\varphi$ can
be written as $r^2\varphi/r^2$, and is thus mildly singular if
$\varphi$ is $C_O$--smooth. In particular, all $h_{ab}$'s and
$h^{ab}$'s are mildly singular if the metric $C$ is smooth.

Denoting by ``m.s." the sum of all mildly singular terms that
might occur, one finds
%
\begin{eqnarray*}
 \partial_a \partial_ b h^{ab}
    & = &
        \Theta^i \Theta^j \Theta^k \Theta^\ell
         \partial_{w^k } \partial_{w^\ell}
           \underline{h_{ij}} + \mbox{m.s.}
 \;,
\\
 2 \partial_a   h^{aB{}}{}_{||B}
    & = &
        -2\Theta^i \Theta^j \Theta^k \Theta^\ell
         \partial_{w^i } \partial_{w^j}
           \underline{h_{ij}}
        -\frac{2n}{r }  \Theta^i \Theta^j \Theta^k
         \partial_{w^k }
           \underline{h_{ij}}
\\
 &&
         +\frac{2n}{r^2} \Theta^i \Theta^j
           \underline{h_{ij}} + \mbox{m.s.}
 \;,
\\
 \frac{n+3}r   h^{  rB{}}{}_{||B}
    & = &-
 \frac{n+3}r  \Theta^i \Theta^j \Theta^k
         \partial_{w^k }
           \underline{h_{ij}}
        -\frac{n(n+3)}{r^2} \Theta^i \Theta^j
           \underline{h_{ij}} + \mbox{m.s.}
 \;,
\\
 \frac{n+1}r \partial_a   h^{a r}
    & = &
        \frac{n+1}{r }  \Theta^i \Theta^j \Theta^k
         \partial_{w^k }
           \underline{h_{ij}}  + \mbox{m.s.}
 \;,
\\
 -\frac{1}r (\partial_r H + \partial_u H)
    & = &
        \frac{1}{r }  \Theta^i \Theta^j \Theta^k
         \partial_{w^k }
           \underline{h_{ij}}  + \mbox{m.s.}
 \;,
\\
 \frac{n-1}{r^2} h^{rr} - \frac 1{r^2} H
    & = &
        \frac{n}{r^2 }  \Theta^i \Theta^j
           \underline{h_{ij}}  + \mbox{m.s.}
 \;.
\end{eqnarray*}
We conclude that
\begin{eqnarray}
 \nonumber
  {h^{AB}{_{||AB}}}
  &= &
   -
    \partial_{w^i} \partial_{w^j} \underline{h_{k\ell }} \Theta^i \Theta^j \Theta^k \Theta^\ell
   -
   \frac {(2n+1)} {r } \partial_{w^i} \underline{h_{jk }} \Theta^i \Theta^j \Theta^k
\\
 &&
 -
    \frac {n^2}{r^2} \underline{h_{ij }} \Theta^i \Theta^j + \mbox{m.s.}
     \;.
 \label{14IX0.6}
\end{eqnarray}
We emphasize that this formula is independent of the ``gauge
condition" $\underline{h_{\mu\nu}} w^\mu=0$.

We now assume that the space dimension $n$ equals three. From
\eqref{13IX0.4} we find
\begin{equation}
 \label{17IX0.2}
 \gamma = \overline{\frac H2} = \frac 12 \overline{\left(
  \eta^{\mu\nu}h_{\mu\nu} +\underline{h_{00}}
           -\underline{h_{ij}}\Theta^{i}\Theta^{j}
           \right)}
 \;,
\end{equation}
which is mildly singular. Let $\chi_{AB}$ denote the $\mathring
s$--trace-free part of $h_{AB}$, then
$$
   {h^{AB}{_{||AB}}} =
   {\chi^{AB}{_{||AB}}} + \frac 1 {r^2} \mathring \Delta \gamma
   \;,
$$
where $\mathring \Delta$ is the Laplace-Beltrami operator of
$\mathring s$. With some work, using $\mathring \Delta \Theta^i
= - 2\Theta^i$, we find
%
\begin{eqnarray}
 \nonumber
 \frac 1 {r^2} \mathring \Delta \gamma
  &= &
   - \frac 12
    \partial_{w^i} \partial_{w^j} \underline{h_{k\ell }} \Theta^i \Theta^j \Theta^k \Theta^\ell
   -
   \frac {3} {r } \partial_{w^i} \underline{h_{jk }} \Theta^i \Theta^j \Theta^k
   \\
 &&
 - \frac {3}{r^2} \underline{h_{ij }} \Theta^i \Theta^j + \mbox{m.s.}
     \;,
 \label{14IX0.7}
\end{eqnarray}
which shows  that $ {\chi^{AB}{_{||AB}}}$ is again of the
general form \eqref{14IX0.6}:
\begin{eqnarray}
 \nonumber
  {\chi^{AB}{_{||AB}}}
  &= &
   - \frac 12
    \partial_{w^i} \partial_{w^j} \underline{h_{k\ell }} \Theta^i \Theta^j \Theta^k \Theta^\ell
   -
   \frac {4} {r } \partial_{w^i} \underline{h_{jk }} \Theta^i \Theta^j \Theta^k
\\
 &&
 -
    \frac {6}{r^2} \underline{h_{ij }} \Theta^i \Theta^j + \mbox{m.s.}
     \;.
 \label{14IX0.9}
\end{eqnarray}

It turns out that things improve  when the normal coordinates
condition is invoked. For then we have
\begin{eqnarray}
 \underline{h_{ij}}  w^i
  & = &
   -
 \underline{h_{0j}}  w^0
  \;,
\\
 \underline{h_{0j}}  w^i
  & = &
   -
 \underline{h_{00}}  w^0
  \;,
\\
 \label{17IX0.6}
 \underline{h_{ij}}  w^i w^j
  & = &
 \underline{h_{00}} ( w^0)^2
  \;,
\\
  w^k  w^i w^j\partial_k \underline{h_{ij}}
  & = &  \big( -2 \underline{h_{00}}
  +
   w^k\partial_k \underline{h_{00}}\big) ( w^0)^2
  \;,
\\
  w^\ell w^k  w^i w^j\partial_\ell \partial_k \underline{h_{ij}}
  & = & \left[ w^\ell \partial_ \ell \big( -2 \underline{h_{00}}
  +
    w^k \partial_k \underline{h_{00}}\big) +3\big( 2 \underline{h_{00}}
  -
   w^k\partial_k \underline{h_{00}}\big)\right]( w^0)^2
  \;.
   \nonumber
\end{eqnarray}
On the light cone this gives
\begin{eqnarray}
 \overline{
 \underline{h_{ij}}
 }
  \Theta^i
  & = &
   -
 \overline{
 \underline{h_{0j}}
 }
  \;,
\\
 \overline{
 \underline{h_{0j}}
 }\Theta^i
  & = &
   -
 \overline{
 \underline{h_{00}}
 }
  \;,
\\
 \label{17IX0.5}
 \frac 1 {r^2}
 \overline{
 \underline{h_{ij}}
 } \Theta^i  \Theta^j
  & = &
 \frac 1 {r^2}
 \overline{
 \underline{h_{00}}
 }
  \;,
\\
 \frac 1 {r }
   \Theta^k   \Theta^i  \Theta^j
 \overline{\partial_k \underline{h_{ij}}
 }
  & = &   -\frac 2{r^2}
 \overline{\underline{h_{00}}
 }
  +
    \Theta^k
 \overline{\partial_k \underline{h_{00}}
 }
  \;,
\\
   \Theta^\ell  \Theta^k   \Theta^i  \Theta^j
 \overline{\partial_\ell \partial_k \underline{h_{ij}}}
  & = &  \frac 1 {r^2}\overline{
    w^\ell
 \partial_ \ell \big( -2 \underline{h_{00}}
  +
   w^k
   \partial_k \underline{h_{00}}\big) +3\big( 2 \underline{h_{00}}
  -
   w^k\partial_k \underline{h_{00}}\big)
   }
  \;.
  \nonumber
\\
 \label{17IX0.4}
\end{eqnarray}
Since all the right-hand sides are mildly singular, from
\eqref{14IX0.9} we conclude that
\begin{eqnarray}
  \mathring \Delta (\mathring \Delta +2) \alpha  &= &
  r^{-2} \mathring s^{AC} \mathring s^{BD} \chi_{CD}{_{||AB}}
 = r^2
 \eta^{AC} \eta^{BD} \chi_{CD}{_{||AB}}
 =
 r^{2} \nonumber {\chi^{AB}{_{||AB}}}
\\
  &= & r^2 \times \, \mbox{m.s.} \;;
 \label{14IX0.10}
\end{eqnarray}
equivalently,
\begin{eqnarray}
 \nonumber
  \mathring \Delta (\mathring \Delta +2) \alpha
 \ \mbox{is $C_O$--smooth.}
 \label{14IX0.10x}
\end{eqnarray}
Up to an element of the kernel of $\mathring \Delta (\mathring
\Delta +2)$, which is irrelevant as it does not contribute to
\eqref{13IX0.4}, we find that $\alpha$ is $C_O$--smooth:
Indeed, if we let $\Pi$ denote the projector, at fixed $r$, on
the space orthogonal to $\ell=0$ and $\ell=1$ spherical
harmonics, we have

\begin{proposition}
 \label{P22IX0.1}
Let $k\in {\mathbf N}\cup \{\infty\}\cup \{\omega\}$, and let
$\mathring\Delta (\mathring \Delta +2)\alpha \in C^k(C_O)$.
Then
$$
 \Pi\alpha\in C^k(C_O)
 \;.
$$
\end{proposition}

\noindent{\sc Proof:}
Assume, first, that $k<\infty$. Let
\begin{equation}
 \label{23XI.1xx}
 \sum_{p=2}^{k}\big(
f_{i_{1}\cdots i_{p}}\Theta ^{i_{1}}\cdots \Theta ^{i_{p}}
+
f^{\prime}{}_{i_{1}\cdots i_{p-1}}\Theta ^{i_{1}}\cdots
\Theta ^{i_{p-1}}\big)r^{p}+o_k(r^{k})
\end{equation}
be the Taylor series of  $\mathring\Delta (\mathring \Delta
+2)\alpha$, as guaranteed by Lemma~\ref{L23I.1}. (The fact that
the series starts at $p=2$ will be justified shortly.)
Decomposing the coefficients $f_{i_{1}\ldots i_{p}}$ and $
f^{\prime}{}_{i_{1}\ldots i_{p-1}}$ into trace terms and
trace-free parts, and rearranging the result, we can without
loss of generality assume that the $f_{i_{1}\ldots i_{p}}$'s
and $ f^{\prime}{}_{i_{1}\ldots i_{p-1}}$'s are traceless. It
then follows from~\cite[pp.~201-202]{GallotHulinLafontaine}
that the finite sums
\begin{equation}
 \label{23XI.1jj}
 \sum_{p \ \mbox{\scriptsize fixed}}
f_{i_{1}\ldots i_{p}}\Theta ^{i_{1}}\cdots \Theta ^{i_{p}}
 \ \mbox{and} \
  \sum_{p \ \mbox{\scriptsize fixed}}
   f^{\prime}{}_{i_{1}\ldots i_{p-1}}\Theta ^{i_{1}}\cdots \Theta
^{i_{p-1}}
\end{equation}
are linear combinations of $\ell=p$, respectively $\ell=p-1$,
spherical harmonics. (This explains why the sum in
\eqref{23XI.1xx} starts with $p=2$, as the image of
$\mathring\Delta (\mathring \Delta +2) $ is orthogonal to
$\ell=0$ and $\ell=1$ spherical harmonics.) Set
\begin{eqnarray*}
\lefteqn{
  \varphi:=  \alpha    - \sum_{p=2}^{k}\frac{1}{p(p+1)(p+2)}
  }
  &&
\\
 &&
   \times\big[
 \frac{1}{(p+3)}f_{i_{1}\ldots i_{p}}\Theta ^{i_{1}}\cdots \Theta ^{i_{p}}
  +
 \frac{1}{(p-1)}f^{\prime}{}_{i_{1}\ldots i_{p-1}}
 \Theta ^{i_{1}}\cdots \Theta ^{i_{p-1}}\big]r^{p}
  \;.
\end{eqnarray*}
Then
\begin{eqnarray*}
 \mathring\Delta (\mathring \Delta +2) \varphi =o_k(r^{k})
 \;.
\end{eqnarray*}
Standard elliptic estimates imply that
$$
 \forall \ 0 \le i \le k \quad \|\partial^i_r \Pi\varphi\|_{H^{k-i}(S^2)} = o(r^{k-i})
 \;,
$$
and our claim easily follows.

If $k=\omega$, convergence for small $|w^0|+ |\vec w|$ of the
series
\begin{equation}
 \label{22IX0.2}
\sum_{p=2} \frac{1}{p(p+1)(p+2)}\big[
 \frac{1}{(p+3)}f_{i_{1}\ldots i_{p}}w^{i_{1}}\cdots w^{i_{p}}
  + w^0
 \frac{1}{(p-1)}f^{\prime}{}_{i_{1}\ldots i_{p-1}}w ^{i_{1}}\cdots
 w^{i_{p-1}} \big]
\end{equation}
follows immediately from that of
$$
\sum_{p=2}  \big(
  f_{i_{1}\ldots i_{p}}\Theta ^{i_{1}}\cdots \Theta ^{i_{p}}
  +
  f^{\prime}{}_{i_{1}\ldots i_{p-1}}\Theta ^{i_{1}}\cdots \Theta ^{i_{p-1}}\big)r^{p}
  \;.
$$

If $k=\infty$ we let $\tilde \alpha$ denote the Borel sum, as
in Appendix~\ref{A19IX0.2}, associated with  \eqref{22IX0.2}.
Then
\begin{eqnarray*}
\forall k\qquad  \mathring\Delta (\mathring \Delta +2) (\alpha-\tilde \alpha) =o_k(r^{k})
 \;,
\end{eqnarray*}
and one concludes as before.
\hfill$\Box$

\bigskip

Returning to our main argument, note that it follows from
\eqref{17IX0.2} and \eqref{17IX0.5} that
\begin{equation}
 \label{17IX0.3}
 \gamma  =  \frac 12 \overline{
  \eta^{\mu\nu}h_{\mu\nu} }
 \;,
\end{equation}
which shows that $\gamma$ is $C_O$--smooth.

We pass now to the term
$$
  \zD_\rho\left( \eta^{\sigma \rho} T_\alpha w_\beta
 \epsilon^{\alpha \beta\gamma\delta} \zD_\gamma h_{\delta\sigma}
 \right)
 \;.
$$
%
%
%
Let $\epsilon_{\mu\nu\rho\sigma}$ be the unique anti-symmetric
tensor such that
$$
 \underline{\epsilon_{0 1 2 3 }} =
 1
 \;, \quad \mbox {we set} \ \epsilon^{AB}= \frac{ w^i}r \underline{\epsilon^{0ijk}}
  \frac{\partial x^A}{\partial w^j}
  \frac{\partial x^B}{\partial w^k}
 \;.
$$
Here, and in what follows, we use the summation convention on
any repeated indices, regardless of their positions. We have
\[  T_\alpha w_\beta
 \epsilon^{\alpha \beta\gamma\delta} \zD_\gamma h_{\delta\sigma}
 =  - w^i
 \epsilon^{0ijk} \zD_j h_{k\sigma}
 = r \epsilon^{AB}\zD_B h_{A\sigma}= r \epsilon^{AB} h_{A\sigma;B} \, , \]
\[ h_{AC;B}= h_{AC||B}+\frac1r\eta_{AB}(h_{uC}+h_{rC})
+\frac1r\eta_{BC}(h_{uA}+h_{rA}) \, ,\]
\[ h_{Aa;B}= h_{Aa||B} + \frac1r\eta_{AB}(h_{ua}+h_{ra}) - \delta^r_a\frac1r h_{AB} \]
\[ \epsilon^{AB} h_{AC;B}= \epsilon^{AB} h_{AC||B} +\frac1r \epsilon^A{_C}
(h_{uA}+h_{rA}) \, ,\]
\[
 \epsilon^{AB} h_{Aa;B}= \epsilon^{AB} h_{Aa||B}\, ,
\]
and finally
\begin{eqnarray}
 \nonumber
 \lefteqn{
 \zD_\rho \left(\eta^{\sigma \rho} T_\alpha w_\beta
 \epsilon^{\alpha \beta\gamma\delta} \zD_\gamma h_{\delta\sigma}\right)
 = \left( r \epsilon^{AB} h_{A\sigma;B}\right)^{;\sigma}
 }
 &&
 \nonumber
\\
 & = &
 \frac1{r^2} \left( r^3 \eta^{ab}\epsilon^{AB} h_{Ab;B}\right)_{,a}
 + \left( r \eta^{CD}\epsilon^{AB} h_{AC;B}\right)_{||D}
 \nonumber
\\
 & = &
 r \epsilon^{AB} \chi_A{^C}{_{||BC}}
 + r \epsilon^{AB}\partial_u \underbrace{ h^u{_{A||B}}}_{ h{_{rA||B}} }
  +
  \frac1{r^3}\partial_r\big( r^4 \epsilon^{AB} \hspace{-0.7cm}
  \underbrace{h^r{_{A||B}}}_{\phantom{X}h{_{uA||B}+h{_{rA||B}}}}
   \hspace{-0.6cm}
  \big)
   \;,
   \label{17IX0.1}
\end{eqnarray}
where, as before, $\chi_{AB}$ is the traceless part of
$h_{AB}$.

The left-hand side of the last equation is a smooth function on
space-time. Next,
\[ \hspace*{-0.05cm}
r \epsilon^{AB} h_{A r ;B} \!=\! \underline{ T_\alpha w_\beta
 \epsilon^{\alpha \beta\gamma\delta} \zD_\gamma h_{\delta\sigma} } \, dw^\sigma(\partial_r)
  \!=\! \underline{ T_\alpha w_\beta
 \epsilon^{\alpha \beta\gamma\delta} \zD_\gamma h_{\delta 0} }+\frac {w^i} r \underline{ T_\alpha w_\beta
 \epsilon^{\alpha \beta\gamma\delta} \zD_\gamma h_{\delta i} }
  \, ,
\]
where the right-hand side is the sum of a smooth function and
of a smooth function divided by $r$. Hence so is its
$\partial_u = - \partial_{w^0}$--derivative, which is the
second term in the last line of \eqref{17IX0.1}. We note the
identity,
\[   r \epsilon^{AB} h_{A u ;B} = - \underline{ T_\alpha w_\beta
 \epsilon^{\alpha \beta\gamma\delta} \zD_\gamma h_{\delta 0} }
  \;,
\]
where the right-hand side is a smooth function on space-time.
We conclude that
\begin{equation}
 \label{12IX0.3}
  \mbox{$ \epsilon^{AB} \chi_A{^C}{_{||BC}}$ is mildly singular.}
\end{equation}
This implies that
\begin{eqnarray}
   \mathring \Delta (\mathring \Delta +2) \beta
    & = &
    \overline{
        r^{-2} \mathring \epsilon^{AB} \mathring s^{ CD} \chi_{AD||BC}
         }
    \nonumber
\\
 \nonumber
 & = &
    \overline{
    r^2 \epsilon^{AB} \eta^{ CD} \chi_{AD||BC}
     }%
\\
 \nonumber
 & = &
    \overline{
    r^2 \epsilon^{AB} \chi_A{^C}{_{||BC}}
    }
\\
 & = &
     r^2 \times
    \, \mbox{m.s.}
 \label{12IX0.4}
\end{eqnarray}
Up to an element of the kernel of $\mathring \Delta (\mathring
\Delta +2)$, which is irrelevant as it does not contribute to
\eqref{13IX0.4}, we find that $\beta$ is $C_O$--smooth. We have
therefore proved necessity in Theorem~\ref{T18IX0.1}.

We wish to show, now, that the conditions of our statement are
sufficient: $C_O$--smooth functions $\alpha$, $\beta$, and
$\gamma$ lead to smooth metrics in normal coordinates. For
this, it is convenient to view tensors on $S^2$ as tensors on
$\mathbf{R}^3$ which are orthogonal to $y^i$ in all indices.
For example, the metric $\mathring s=\mathring s_{AB} dx^A
dx^B$ is identified with $r^{-2}$ times the projector
$$P_{ij} = \delta_{ij}
- \frac{w^ i w^j}{ r^2}
 \;.
$$
Indeed,
$$
\mathring s_{AB} dx^A
dx^B = \mathring s_{AB} \frac{ \partial x^A}{\partial w^i}
 \frac{ \partial x^B}{\partial w^j}  dw^i dw^j = r^{-2} \left(\delta_{ij}
- \frac{w^ i w^j}{ r^2}\right) dw^i dw^j
 \;.
$$
 So, if $Y_i$ or
$S_{ij}$ are tensors satisfying $Y_i w^i = 0 = S_{ij} w^j =
S_{ij} w^i$, we have the formulae
$$
 {\mathcal D}_i Y_j = P_i {}^kP_j {}^\ell \partial_k Y_\ell\;,
 \quad
 {\mathcal D}_i S_{jm} = P_i {}^kP_j {}^\ell P_m {}^n \partial_k S_{\ell n}\;,
 \quad
 {\mathcal D}_i S^i{}_{ m} = P_\ell {}^k  \partial_k S_{\ell n}
 \;.
$$
In this formalism we have ${\mathcal D}_i f = P_i{}^j \partial
_j f$, and
\begin{eqnarray*}
{\mathcal D}_i {\mathcal D}_j f  &= &  P_i {}^kP_j {}^\ell
\partial_k (P_\ell{}^m \partial_m f)
= P_i {}^kP_j {}^\ell
\partial_k  \partial_\ell f - \frac 1 r P_{ij} \Theta^m \partial_ m f
 \;.
\end{eqnarray*}
Hence
\begin{equation}
P^{ij} {\cal D}_i {\cal D}_j f = P^{ij} \partial_ i \partial_j f - \frac{ 2}{r} \Theta^m \partial_m f
 \;.
 \label{17IX0.7}
\end{equation}
Let us write
$$
 \alpha = \alp+ r \alt
 \;,
$$
where $\alp$ and $\alt$ are smooth functions of $\vec w$. We
note that
$$
  {\cal D}_i {\cal D}_j \alpha =
  {\cal D}_i {\cal D}_j \alp
  +r {\cal D}_i {\cal D}_j \alt
 \;.
$$
Equation \eqref{17IX0.7} with $f$ replaced by $\alp$ gives
\begin{eqnarray}
 \nonumber
 \lefteqn{
 r^2 \big( 2 {\cal D}_i {\cal D}_j \alp  -P_{ij}P^{k\ell} {\cal D}_k {\cal D}_\ell\alp \big)
  }
  &&
\\
 \nonumber
 & =
  & r^2 \big(2 P_i {}^kP_j {}^\ell
\partial_k  \partial_\ell  \alp- P_{ij}P^{k\ell} \partial_k \partial_\ell\alp
\big)
\\
 & =
  & \Theta^i \Theta^j w {}^k w^\ell \partial_k  \partial_\ell  \alp
  - 2 w^i w^\ell \partial_j  \partial_\ell  \alp
  - 2 w^j w^\ell \partial_i  \partial_\ell  \alp
 \nonumber
\\
 &   &
  - (r^2\delta_i^j - w^i w^j)  \partial_\ell \partial_\ell\alp
 \;.
\end{eqnarray}
An identical calculation applies to $\alt$. We conclude that
the tensor field \eqref{18IX0.1} contains $\Theta \otimes
\Theta$ terms of the form as in \eqref{14IX0.1}, with
\begin{equation}
 \label{29IX0.1}
 A_{ij} = \partial_i \partial_ j \alp
 +r \partial_i \partial_ j \alt
 \;.
\end{equation}

Next, we write
$$
 \beta=\betar+r\betat
 \;,
$$
where $\betar$ and $\betat$ are smooth functions of $\vec w$.
The contribution of $\betar$ to the tensor field
\eqref{18IX0.1} can be rewritten as
\begin{eqnarray*}
&&
 rw^\ell \left( {\epsilon_{k\ell i} \partial_{w^{j}}\partial_{w^k} \betar} +
  { \epsilon_{k\ell j} \partial_{w^{i}}\partial_{w^k} \betar}\right)
\\
&&
 + w^\ell w^m \left( \Theta^{i}\epsilon_{j k\ell } + \Theta^{j}\epsilon_{i k\ell }\right)
   \partial_{w^{k}}\partial_{w^m} \betar
 \;,
\end{eqnarray*}
with a similar formula for $\betat$. The resulting $\Theta$
terms are of the right-form
\begin{eqnarray}
 \nonumber
   w^ mw^\ell\big(
 \underline{{A_{\ell i m}}}
 \Theta^j
 +
 \underline{{A_{\ell j m}}}    \Theta^i
  \big)
\end{eqnarray}
as in \eqref{14IX0.1} if we set
\begin{equation}
 \label{29IX0.2}
 A_{\ell j m} = \epsilon_{ m  k  \ell}( \partial_{w^j} \partial_{w^k} \betar + r \partial_{w^j} \partial_{w^k} \betat)
 - \epsilon_{ j  k  \ell}( \partial_{w^m} \partial_{w^k} \betar + r \partial_{w^m} \partial_{w^k} \betat)
 \;.
\end{equation}

To summarize: let $\Omega_{0i0j}$ be a smooth extension of
$A_{ij}$ as given by \eqref{28IX0.1}, and let $\Omega_{0ijk}$
be a smooth extension of $A_{ijk}$ as given by \eqref{29IX0.2},
if we set $\Omega_{ijkl}=0$, then the restrictions to the
light cone of the $ij$ components of the tensor field
$$
 (\underline{\eta_{\mu\nu}} + \underline{\Omega_{\mu\alpha\rho\beta}} w^\alpha w^\beta) dw^\mu dw^\nu
$$
reproduce the non-manifestly $C_O$--smooth terms in
\begin{eqnarray*}
  r^2 \big[(1+ \gamma)\mathring s_{AB}+ 2\alpha_{||AB}
  - \mathring s_{AB} \mathring s^{CD} \alpha _{||CD}
  + \mathring \epsilon_A{}^C  \beta_{||CB}
  + \mathring \epsilon_B{}^C  \beta_{||CA}\big]
   \frac{\partial x^A}{\partial w^i}
   \frac{\partial x^B}{\partial w^j}
  \;   .
\end{eqnarray*}
So the difference is a $C_O$--smooth tensor field, say $f_{ij}
= \hf_{ij} + r\cf_{ij}$, with $\hf_{ij}$ and $\cf_{ij}$ smooth
tensors on $\R^3$, that satisfies
\begin{equation}
 \label{19IX0.1}
 f_{ij} w^j =0
 \;.
\end{equation}
Now, it is not directly apparent that we have the desired
formula, as in Proposition~\ref{P10IX0.1},
\begin{equation}
 \label{19IX0.2}
 \underline{f_{ij}} = \underline{ A_{ik j\ell} } \, w^k w^\ell
\end{equation}
for some tensor field $A_{ijkl}$ with the right symmetries,
because $f_{ij}$ is not differentiable. However, one can
proceed as follows: Let $\hf_{ij k_1\ldots k_\ell}$ be the
Taylor expansion coefficients of $\hf$,
$$
 \forall m \qquad \hf_{ij} (\vec w)= \sum_{0\le \ell\le m} \hf_{ij k_1\ldots k_\ell} w^{k_1}\cdots w^{k_\ell} + o_m(r^m)
 \;,
$$
similarly for  $\cf_{ij k_1\ldots k_\ell}$. Then the
coefficients in the Taylor expansion of $f_{ij}w^i$ have to
vanish at every power of $r$, which implies that for all
$\ell\in \mathbf N$ we have
$$
 \sum_{\mbox{\scriptsize fixed $ \ell$}}
  \big( \hf_{ij k_1\ldots k_\ell} \Theta^{k_1}\cdots \Theta^{k_{\ell}} +
 \cf_{ij k_1\ldots k_{\ell-1}} \Theta^{k_1}\cdots\Theta^{k_{\ell-1}}\big)\Theta^i
  =0
   \;.
$$
Equivalently,
$$
 \sum_{\mbox{\scriptsize fixed $ \ell$}}
  \big( \hf_{ij k_1\ldots k_\ell} w^{k_1}\cdots w^{k_{\ell}} + r
 \cf_{ij k_1\ldots k_{\ell-1}} w^{k_1}\cdots w^{k_{\ell-1}}\big)w ^i
  =0
   \;.
$$
Comparing this equation with the equation where $w^k$ is
replaced by $-w^k$ we easily conclude that
$$
 \hf_{i(j k_1\ldots k_\ell)} =0=\cf_{i(j k_1\ldots k_\ell)}
 \;.
$$
Let $\widetilde{\cf}_{ij}$ be obtained by Borel summation of
the Taylor series of $\cf_{ij}$, as in Appendix~\ref{A19IX0.2}.
Then each partial sum  $(\widetilde{\cf}_{ij})_p$ as defined in
\eqref{19IX0.5} has vanishing contraction with $w^i$, and so
$\widetilde{\cf}_{ij}w^i=0$ as well by passing to the limit.
Since $\cf_{ij}$ and $\widetilde{\cf}_{ij}$ have the same
Taylor coefficients it holds that
$$
 \forall \ m \qquad {\cf}_{ij}-\widetilde{\cf}_{ij} = o_m(r^m)\;,
$$
where we write $\psi=o_m(r^m)$ if $\psi$ is $m$--times
differentiable with \\ $\displaystyle\lim_{r\to0} \partial_{k_1}\cdots
\partial_{k_\ell} \psi=0$ for $0\le \ell \le m$.  This implies that
$r({\cf}_{ij}-\widetilde{\cf}_{ij})$ is smooth. Hence
$$
 \hf_{ij}+r(
 {\cf}_{ij}-\widetilde{\cf}_{ij} )
$$
is a smooth tensor field satisfying
$$
 \big[\hf_{ij}+r({\cf}_{ij}-\widetilde{\cf}_{ij})\big]w^i =0
 \;.
$$
By Proposition \ref{P10IX0.1} we can write
$$
 \hf_{ij}+r({\cf}_{ij}-\widetilde{\cf}_{ij}) = \hat A_{ikj\ell} w^k w^\ell\;, \quad \widetilde{\cf}_{ij}   = \check A_{ikj\ell} w^k w^\ell
 \;.
$$
This shows that
$$f_{ij}  = \big(\underbrace{\hat A_{ikj\ell}+ r \check A_{ikj\ell}}_{=:A_{ikj\ell}}\big) w^k
w^\ell
 \;,
$$
as desired.

One concludes using Proposition~\ref{P14IX0.1}.
\hfill$\Box$

\section{Other adapted coordinate systems}
 \label{s22IX0.1}

So far we have concentrated on normal coordinates, as these are
naturally singled out by the geometry. However, other (local)
coordinate systems $y^\mu$ in which $C_O$ takes the standard
form $\{y^0 = |\vec y|\}$ exist, and can be useful for some
purposes. The simplest possibility is provided by coordinate
systems of the form
\begin{equation}
 \label{22IX0.3}
 y^\mu = w^\mu + \underline{\eta_{\alpha\beta}} w^\alpha w^\beta \chi^\mu
 \;,
\end{equation}
for some smooth functions $\chi^\mu$. It is likely that all
coordinate systems for which $C_O=\{y^0 = |\vec y|\}$ are
related to the normal ones in this way, but we are not aware of
a proof of this except in the analytic case in dimension $3+1$.

For sufficiently small $|w^0| + |\vec w|$ the inverse
transformation to \eqref{22IX0.2} takes a similar form
\begin{equation}
 \label{22IX0.1}
 w^\mu = y^\mu + \underline{\eta_{\alpha\beta}}y^\alpha y^\beta \psi^\mu
 \;,
\end{equation}
for some smooth functions $\psi^\mu$.

To avoid ambiguities, let us write
$$
 g = g_{y^\mu y^\nu} dy^\mu dy^\nu =
 g_{w^\mu w^\nu} dw^\mu dw^\nu
 \equiv
 \underline{g_{\mu \nu}} dw^\mu dw^\nu
 \;;
$$
one finds
$$
 \overline{ g_{y^\mu y^\nu}}
  =
 \overline{ g_{w^\mu w^\nu}}
  +
  2 \overline{ g_{w^\alpha w^\nu} \chi^\alpha } y_\mu
  +
  2 \overline{ g_{w^\alpha w^\mu} \chi^\alpha } y_\nu
  +
   4 \overline{ g_{w^\alpha w^\beta} \chi^\alpha \chi^\beta} y_\mu y_\nu
   \;,
$$
where $y_ \alpha=\eta_{y^\alpha y^\beta} y^\beta$, with $
\eta_{y^\mu y^\nu}=\mbox{\rm diag}(-1,+1,\ldots,+1)$. Clearly
$\{ \eta_{y^\mu y^\nu} y^\mu y^\nu =0\}$ remains a null
hypersurface on geometric grounds; a useful consistency check
in subsequent calculations is to note that the last equation
implies
\begin{equation}
 \label{23IX0.1}
 \overline{ g_{y^\mu y^\nu}} y^\nu
  =
 \overline{ g_{w^\mu w^\nu}} y^\nu =
  { \eta_{y^\mu y^\nu}} y^\nu
   \;.
\end{equation}
To avoid a proliferation of notation, we will again use the
symbols $x^\alpha$ to denote coordinates defined as
\begin{equation}
y^{0}=x^{1}-x^{0},
\quad
y^{i}=x^1\Theta^{i}(x^{A})
\quad \text{with, as before,} \quad
\sum_{i=1}^{n}\left[\Theta^{i}(x^{A})\right]^{2}=1
 \;.
 \label{22IX0.4}
\end{equation}
 It follows from \eqref{10IX0.7}-\eqref{10IX0.10} that the new
$h_{\mu\nu}=g_{\mu\nu}-\eta_{\mu\nu}$ takes on $C_O$ the form
\begin{eqnarray}
 \label{10IX0.7x}
 \overline{{h_{00}}}
  & = &
 \overline{\underline{\Omega_{0i0j}}} y^i y^j - 4 r
 \overline{\underline{g_{\mu 0} \chi^\mu}}+ 4 r^2
 \overline{\underline{g_{\mu \nu} \chi^\mu\chi^\nu}}
 \;,
\\
 \label{10IX0.8x}
 \overline{{h_{0 1}}}
  & = & 2 \big(r
 \overline{\underline{g_{\mu 0} \chi^\mu}}
 -
 \overline{\underline{g_{\mu i} \chi^\mu}} y^i
  \big)
   \;,
\\
 \label{10IX0.8xx}
 \overline{{h_{1 A}}}
  & = &
 \overline{{h_{1 1}}}
  =
   0
   \;,
\\
 \label{10IX0.9x}
 \overline{{h_{0A}}}
  & = & \bigg[r\big(
 \overline{\underline{\Omega_{0j0i}}} r
  -
 \overline{\underline{\Omega_{0jik}}} y^k \big) y^j
 + 2 r^2 \overline{\underline{g_{\mu i} \chi^\mu}}
 \bigg]
 \frac{\partial \Theta^i}{\partial x^A}
 \;,
\\
 \overline{{h_{AB}}}
 & = &
 r^2
  \bigg[
 \overline{\underline{\Omega_{i0j0}}}r^2
 \nonumber
\\
 \label{10IX0.10x}
  &&
   +
   \big(
   -
 \overline{\underline{\Omega_{0ijk}}}r
  -
 \overline{\underline{\Omega_{0jik}}}r
  +
 \overline{\underline{\Omega_{ikj\ell}}} y^\ell
  \big)y^k
   \bigg]
   \frac{\partial \Theta^i}{\partial x^A}
   \frac{\partial \Theta^j}{\partial x^B}
 \;.
\end{eqnarray}
%

\appendix

\section{Extending functions}
 \label{A22IX0.1}

\begin{lemma}
 \label{L23I.1}
A function $\varphi$ defined on a light cone $C_{O}$ is the
trace $\overline{f}$ on $C_{O}$ of a $C^{k}$ spacetime function
$f$ if and only if $\varphi$ admits an expansion, for small
$r$, of the form
\begin{equation}
 \label{23XI.1}
\varphi= \sum_{p=0}^{k}f_{p}r^{p}+o_k(r^{k})
 \;,
\end{equation}
with
\begin{equation}
 \label{6XI.1}
f_{p}\equiv
f_{i_{1}\ldots i_{p}}\Theta ^{i_{1}}\cdots \Theta ^{i_{p}}
+
f^{\prime}{}_{i_{1}\ldots i_{p-1}}\Theta ^{i_{1}}\cdots \Theta ^{i_{p-1}}
 \;,
\end{equation}
where $f_{i_{1}\ldots i_{p}}$ and $f_{i_{1}\ldots
i_{p-1}}^{\prime }$ are numbers.

The claim remains true with $k=\infty$ if \eqref{23XI.1} holds
for all $k$.
\end{lemma}

{\noindent\sc Proof:} The result is trivial away from the
origin, so it suffices to consider functions defined near the
tip of the light cone.

{Suppose, first, that $k<\infty$.} To see the necessity, let
$f$ be a function which is $C^{k}$ in a neighbourhood of the
origin in $\mathbf{R}^{n+1}$. For any multi-index
$\beta=(\beta_1,\ldots, \beta_j)\in (\mathbf{N}^{n+1})^{j }$,
$\beta_i \in \{0,1,\ldots , n\}$,  with length $1\le |\beta|:=j
\le k$ set
$$
 f_\beta :=  \frac{\partial}{\partial y^{\beta_1}}  \cdots  \frac{\partial}{\partial y^{\beta_j}} f
  \;.
 $$
Then $f_\beta$ is $C^{k-|\beta|}$ in a neighbourhood of the
origin, and thus admits a Taylor expansion
\begin{equation}
f_\beta =\underbrace{\sum_{p=0}^{k-|\beta|}h_{\beta;\alpha _{1}\cdots \alpha _{p}}y^{\alpha
_{1}}\cdots y^{\alpha _{p}}}_{=:h_\beta }+\underbrace{g_\beta  }_{o(|y|^{k-|\beta|})}
 \;,
\end{equation}
for some coefficients $h_{\beta;\alpha _{1}\cdots \alpha
_{p}}\in \mathbf{R}$. Since $f_\beta \in C^{k-|\beta|}$ and
$h_\beta \in C^\infty$ we have $g_\beta=f_\beta - h_\beta \in
C^{k-|\beta|}$. Similarly
\begin{equation}
f  = \underbrace{\sum_{p=0}^{k}f_{\alpha _{1}\cdots \alpha _{p}}y^{\alpha
_{1}}\cdots y^{\alpha _{p}}}_{=:h }+\underbrace{g}_{o(|y|^{k})}
 \;,
\end{equation}
with $f_{\alpha _{1}\cdots \alpha _{p}} \in \mathbf{R}$, $h \in
C^\infty$ and $g\in C^{k}$. The usual formula for the
coefficients of a Taylor expansion implies that
$$
 h_\beta  = \frac{\partial}{\partial y^{\beta_1}}  \cdots  \frac{\partial}{\partial y^{\beta_j}} h
  \;.
 $$
Hence
\begin{eqnarray}
 \nonumber
 \frac{\partial}{\partial y^{\beta_1}}  \cdots  \frac{\partial}{\partial y^{\beta_j}} g
  & = &
    \frac{\partial}{\partial y^{\beta_1}}  \cdots  \frac{\partial}{\partial y^{\beta_j}}(f-h)
\\
 & = &
 f_\beta -h_\beta  = g_\beta   = o(|y|^{k-j})
 \;,
  \label{14I.1}
\end{eqnarray}
and so $g=o_k(|y|^k)$. Now, $\overline{f} =
\overline{h}+\overline{g}$, and it should be clear that
$\overline h$ is of the form (\ref{6XI.1}). The estimate
$\overline g=o_k(r^k)$ is then straightforward from
$g=o_k(|y|^k)$, using
\[
 \frac{\partial}{\partial y^{i_1}} \cdots \frac{\partial}{\partial y^{i_j}} \overline g =
 \overline{ \left( \frac{y^{i_1}}r  \frac{\partial}{\partial y^0} + \frac{\partial}{\partial y^{i_1}}\right)  \cdots
 \left( \frac{y^{i_j}}r \frac{\partial}{\partial y^0}  + \frac{\partial}{\partial y^{i_j}}\right)
 g}
 \;.
\]

Conversely, let $\varphi = \psi + \chi$ be defined on a
neighbourhood of $O$ on $C_O$, where
\begin{eqnarray*}
\psi
 &= &  \sum_{p=0}^{k}(f_{i_{1}\ldots i_{p}}\Theta ^{i_{1}}\cdots \Theta ^{i_{p}}
 +
 f^{\prime}{}_{i_{1}\ldots i_{p-1}}\Theta ^{i_{1}}\cdots \Theta ^{i_{p-1}})r^p
\\
 & = &
\sum_{p=0}^{k}(f_{i_{1}\ldots i_{p}}y ^{i_{1}}\cdots y^{i_{p}}
 +
 r f^{\prime}{}_{i_{1}\ldots i_{p-1}}y ^{i_{1}}\cdots y ^{i_{p-1}})
 \;,
\end{eqnarray*}
and where $\chi = o_k(r^k)$. Set $t =y^0$, $\vec y=(y^1,\ldots ,
y^n)$, and
$$
 f(t,\vec y ) = \sum_{p=0}^{k}(f_{i_{1}\ldots i_{p}}y ^{i_{1}}\cdots y^{i_{p}}
 +
 t f^{\prime}{}_{i_{1}\ldots i_{p-1}}y ^{i_{1}}\cdots y ^{i_{p-1}})
 +\chi
 \;.
 $$
Then $\overline f = \varphi$. The function $\chi(\vec y)$,
viewed as a function of $(t,\vec y)$, is trivially $ o_k
(|y|^k)$, and the proof is completed for finite $k$.

The case $k=\infty$ is obtained from the above by Borel
summation, using Lemma~\ref{LBorel}, Appendix~\ref{A19IX0.2}.
\hfill$\Box$

 \section{How to recognize that coordinates are normal}
  \label{A19IX0.1}

In this appendix we prove some simple necessary and sufficient
conditions for a coordinate system to be normal:

\begin{Proposition}[Thomas~\cite{Thomas}]
\label{Pnsno} Let $\{x^\mu\}$ be a local coordinate system
defined on a star shaped domain containing the origin. The
following conditions are equivalent:
\begin{enumerate}
\item For every $a^\mu \in {\mathbf{R}}^n$ the rays $s\to
    sa^\mu$ are geodesics;
\item $\Gamma^\mu{}_{\alpha\beta}(x)x^\alpha x^\beta = 0 $;
\item $\displaystyle\frac{\partial
    g_{\gamma\alpha}}{\partial x^\beta}(x) x^\alpha x^\beta
    = 0$;
\item \label{Pnsno4}$g_{\alpha\beta}(x)x^\beta
    =g_{\alpha\beta}(0)x^\beta$.
\end{enumerate}
\end{Proposition}

\medskip

\noindent{\sc Proof:} $1.\Leftrightarrow 2.$: The rays
$\gamma^\mu(s)=sa^\mu$ are geodesics if and only if
$$0=\underbrace{\frac{d^2\gamma^\mu}{ds^2}}_{=0}+\Gamma^\mu{}_{\alpha\beta}(sa^\sigma)
\frac{d\gamma^\alpha}{ds} \frac{d\gamma^\beta}{ds}
=\Gamma^\mu{}_{\alpha\beta}(sa^\sigma) a^\alpha a^\beta\;,$$
multiplying by $s^2$ and setting $x^\mu=sa^\mu$ the result
follows.

$3.\Leftrightarrow 4.$: \begin{eqnarray}
g_{\mu\alpha}(x^\sigma)x^\alpha = g_{\mu\alpha}(0)x^\alpha &
\Longleftrightarrow & g_{\mu\alpha}(s a^\sigma)a^\alpha =
g_{\mu\alpha}(0)a ^\alpha
\\ & \Longleftrightarrow & \frac
d{ds}\left(g_{\mu\alpha}(sa^\sigma)a^\alpha\right)=0\\
& \Longleftrightarrow &\frac{\partial
g_{\mu\alpha}(x^\sigma)}{\partial x^\beta} x^\alpha x^\beta
=0\;. \end{eqnarray}

$2.\Rightarrow 4.$: From the formula for the Christoffel
symbols in terms of the metric we have
\begin{equation}\label{equivG}\Gamma^\mu{}_{\alpha\beta}(x)x^\alpha x^\beta = 0
\quad \Longleftrightarrow \quad \left(2 \frac{\partial
g_{\mu\alpha}}{\partial x^\beta} - \frac{\partial
g_{\alpha\beta}}{\partial x^\mu}\right)x^\alpha x^\beta = 0
\;.\end{equation} Multiplying by $x^\mu$ we obtain \begin{eqnarray}
\frac{\partial g_{\mu\alpha}(x^\sigma) }{\partial
x^\beta}x^\alpha x^\beta x^\mu= 0 & \Longleftrightarrow &
\frac{\partial g_{\mu\alpha}(s a^\sigma) }{\partial
x^\beta}a^\alpha a^\beta a^\mu= 0
\\ & \Longleftrightarrow &
\frac {d}{ds}\left(g_{\mu\alpha}(s a^\sigma)a^\alpha
a^\mu\right)=0\\ & \Longleftrightarrow & g_{\mu\alpha}(s
a^\sigma)a^\alpha a^\mu=g_{\mu\alpha}(0)a^\alpha a^\mu \\ &
\Longleftrightarrow & g_{\mu\alpha}(x^\sigma)x^\alpha
x^\mu=g_{\mu\alpha}(0)x^\alpha x^\mu \;.\end{eqnarray}
Differentiating it follows that
$$\frac{\partial g_{\mu\alpha}(x^\sigma)}{\partial x^\gamma} x^\alpha
x^\mu + 2 g_{\gamma\alpha}(x^\sigma)x^\alpha
=2g_{\gamma\alpha}(0)x^\alpha  \;.$$ Substituting this into the
last term in \eqref{equivG} one obtains
\begin{equation}\label{oncag} \frac{\partial
g_{\mu\alpha}}{\partial x^\beta} (x^\sigma) x^\alpha x^\beta
+g_{\mu\alpha}(x^\sigma)x^\alpha -g_{\mu\alpha}(0)x^\alpha
=0\;.\end{equation} This implies that
$$\frac d{ds}\left[g_{\mu\alpha}(sa^\mu)sa^\alpha -
g_{\mu\alpha}(0)sa^\alpha\right]=0\;,$$ and the result follows by
integration.

$3.\&4.\Rightarrow 2.$:  Point 4 implies
$$g_{\alpha\beta}(x^\gamma) x^\alpha x^\beta =
g_{\alpha\beta}(0) x^\alpha x^\beta\;.$$ Differentiating  one
obtains
$$\frac{\partial g_{\alpha\beta}(x^\gamma)}{\partial x^\mu} x^\alpha
x^\beta + 2g_{\alpha\mu}(x^\gamma) x^\alpha  = 2g_{\alpha\mu}(0)
x^\alpha  \;.$$ The last two terms are equal by point 4 so that
$$\frac{\partial g_{\alpha\beta}(x^\gamma)}{\partial x^\mu} x^\alpha
x^\beta =0\;.$$ This shows that the last term in \eqref{equivG}
vanishes, so does the next-to-last by point 3, and the proof
is complete.
\hfill $ \Box$
\newcommand{\jj}[1]{{ {\mnote{{\bf jj:}#1}}}}
\newcommand{\halpha}{\hat \alpha}
\newcommand{\calpha}{\check \alpha}
\newcommand{\hbeta}{\hat \beta}
\newcommand{\cbeta}{\check \beta}
\section{Covector fields}
 \label{A24IX0.1}

The aim of this appendix is to present a simple equivalent of
our parameterization of the metric for covector fields. This
can be used for Cauchy problems on the light cone involving
Maxwell fields.

We start by noting that every covector field $\zeta_\mu$ on
space-time can be written as
$$
 \zeta_\mu = \xi_\mu + \partial_ \mu \lambda\;,
 \
 \mbox{with}
 \
 w^\mu \xi_ \mu =0
 \;,
$$
for a smooth function $\lambda$. This is obtained by setting
$$
 \lambda( w^\mu) = w^\alpha \int_0^1 \zeta_ \alpha( sw^\mu) ds
  \;.
$$
By the arguments in Section~\ref{ss14IX0.1} there exists a
smooth anti-symmetric matrix  $\Omega_{\mu\nu}$ such that
\begin{equation}
 \label{24IX0.2}
 \underline{\xi_ \mu}
  = \underline{\Omega_{\mu\nu}} w^\nu
  \;.
\end{equation}
As in the main body of this paper, the restriction to the
light cone $\{w^0 = |\vec w|\}$ of $\xi_ \mu$ arises from a
smooth vector field on $\R^{4}$ satisfying $w^\mu
\underline{\xi_\mu} =0$ if and only if the restrictions
$\overline{\underline{\Omega_{\mu\nu}}}$ are $C_O$--smooth.

An alternative parameterization of $\xi$ is obtained by
introducing
\begin{eqnarray}
 \xi_u= - \underline{\xi_0}
 \;,
 \quad
  \gamma=\Theta^{i}\overline{\underline{\xi_i}} \;, \quad \xi_A =
 \underline{\xi_i} w^i_{,A}
\;,\quad
 \overline{\xi_A} = \alpha _{||A} +  \epsilon_A{}^C  \beta_{||C}
  \;,
\label{covector}
 \end{eqnarray}
and we have $\overline{\xi_u}=\gamma$ in view of the condition
$\xi_ \mu w^\mu =0$. We then have:

\begin{theorem}
 \label{jjT18IX0.1}
A field  of the form \eqref{covector} defined on $(0,R)\times
S^2$ is the restriction to the light cone $\{w^0 = |\vec w|\}$
of a smooth vector field on $\R^{4}$ satisfying $w^\mu
\underline{\xi_\mu} =0$ if and only if
$$
 \mbox{ $\alpha=r\calpha + r^2 \halpha$, $\beta=r\cbeta  +
r^2 \hbeta    $,  and $\gamma=w^{i}\partial_{w^i}\calpha    +
r \check \gamma +r^2 \hat \gamma    $,
 }
$$
where
$$
 \mbox{ $\calpha$, $\halpha$, $\cbeta$, $ \hbeta    $, $\check \gamma$ and $ \hat \gamma    $ are smooth functions of $\vec w$,
 }
$$
except for the $\ell=0$ spherical-harmonics components of $
{\alpha}$ and ${\beta}$ which do not affect $\xi_ \mu$.
\end{theorem}

\noindent{\sc Proof:}
Necessity: it follows from the identities
\begin{equation}
 \label{27IX0.3}
  r^2 \overline{\xi^A{_{||A}}}= \mathring \Delta \alpha=
\overline{r^2 \partial_{w^k}
 \underline{\xi^k} - w^jw^i \partial_{w^i}\underline{\xi_j}
 -2 r w^i \underline{\xi_i}}
 \;,
\end{equation}
\begin{equation}
 \label{27IX0.4}
 r^2 \overline{\epsilon^{AB} \xi_{A||B}}= \mathring \Delta \beta = \overline{r w_i
 \epsilon^{ijk} \partial_{w^k} \underline{\xi_j}}
 \;,
\end{equation}
together with a straightforward generalization of
Proposition~\ref{P22IX0.1} that $\alpha$ and $r^{-1}\beta$ are
$C_O$--smooth if $\xi_\mu$ is smooth, except for their $\ell=0$
components which are in the kernel of $\mathring \Delta$.
However, the gauge condition $0=\underline{\xi_ \mu} w^\mu =w^0
\underline{ \xi_0 } + w^i \underline {\xi_ i} $ implies
$$ { w^j w^i \partial_j \underline{\xi_i}} =  -t  w^j  {
\partial_j\underline{\xi_0}} + t \underline{\xi_0}
\;,
$$
and we conclude that
\begin{eqnarray}
 \label{27IX0.1}
&
 \overline{  \Theta^i \underline{\xi_i}} = - \overline{
 \underline{\xi_0}} \;,
&
\\
&
 \label{27IX0.2}
  \overline{ w^j w^i \partial_j \underline{\xi_i}} =  -r  w^j  \overline{
\partial_j\underline{\xi_0}} + r \overline{\underline{\xi_0}}
\;.
&
\end{eqnarray}
The $C_O$--smoothness of $\gamma$  follows from
\eqref{27IX0.1}, while that of  $\alpha/r$ follows from
\eqref{27IX0.3} and \eqref{27IX0.2}.

We can write
$$
 \gamma=w^{i}\partial_{w^i}\calpha     + \dgamma
 \;,
$$
and it remains to show that $\dgamma /r$ is $C_O$--smooth.
 The  inverse of \eqref{covector} reads
\begin{eqnarray}
 \nonumber
 \overline{\underline{\xi_k}}
  & = & r\partial_{w^k}\calpha  +
    w^{i}\epsilon_{ik}{^l} \partial_{w^l}\cbeta
 +
     r^2\partial_{w^k}\halpha  +
    r w^{i}\epsilon_{ik}{^l} \partial_{w^l}\hbeta  -w^{i}\partial_{w^i}\halpha w^{k}
\\
 &&
    + \dgamma  \Theta^k
    \;.
 \label{24IX0.1}
\end{eqnarray}
Extending $\halpha$, $\calpha$, etc., to ${\mathbf R}^{4}$ by
requiring the extension to be time-independent, and using the
same symbols for this extension, $\overline{\underline{\xi_k}}$
minus the first line of the right-hand side of \eqref{24IX0.1}
is the restriction to the light cone of the smooth vector field
\begin{eqnarray}
 \nonumber
  {\underline{\xi_k}}
  & - & \big( t\partial_{w^k}\calpha  +
    w^{i}\epsilon_{ik}{^l} \partial_{w^l}\cbeta
\\
 &&
 +
     r^2\partial_{w^k}\halpha  +
    t w^{i}\epsilon_{ik}{^l} \partial_{w^l}\hbeta  -w^{i}\partial_{w^i}\halpha\, w^{k}
    \big)
    \;.
 \label{24IX0.10}
\end{eqnarray}
Hence for every $k$ the function
$$ \dgamma  (\vec w)
 \Theta^k = \frac{\dgamma (\vec w)}r w^k
$$
extends to a smooth function on space-time. Choosing $k$ to be
one, by Proposition~\ref{P18IX0.1} we can write
\begin{equation}
 \label{28IX0.1}
 \frac{\dgamma (\vec w)}r w^1 = \check\chi(\vec w) + r \hat \chi(\vec w)
  \;,
\end{equation}
for some smooth functions $\check \chi$ and $\hat \chi$. For
$r\ne 0$ this implies
$$\left[\check\chi(\vec w) + r \hat \chi(\vec
w)\right]\big|_{w^1 =0} =0\; ;$$ by continuity this holds for all $r$.
Smoothness of $\check\chi$ and $\hat \gamma $ implies existence
of smooth functions $\check\gamma$ and $\hat \gamma  $ such
that
$$
 \check \chi = \check \chi|_{w^1=0} + \check\gamma w^1 \;,
 \quad
 \hat \chi = \hat \chi|_{w^1=0} + \hat \gamma  w^1 \;,
$$
and \eqref{28IX0.1} gives
\begin{equation}
 \label{28IX0.2}
 \frac{\dgamma (\vec w)}r w^1 =
 \big[\check\gamma(\vec w) + r \hat \gamma (\vec w)\big]w^1
  \;.
\end{equation}
For $w^1\ne 0$ we conclude
\begin{equation}
 \label{28IX0.3}
  {\dgamma (\vec w)}  = r\check\gamma(\vec w) + r^2 \hat \gamma (\vec w)
  \;,
\end{equation}
and continuity implies that this equation holds everywhere. We
conclude that $\dgamma /r$ is $C_O$--smooth, and the proof of
necessity is complete.

Sufficiency should be clear from what has been said together
with
\begin{equation}
 \label{28IX0.4}
  {\dgamma (\vec w)} \Theta^k   = \overline{
  \big[\check\gamma(\vec w) + t \hat \gamma (\vec w)\big] w^k}
  \;.
\end{equation}
\hfill $\Box$

\section{Borel's summation}
 \label{A19IX0.2}

In the main body of the paper we will need the details of the
following construction, which is a straightforward adaption
of~\cite[Volume~I, Theorem~1.2.6]{HormanderAPDE}:

\begin{lemma}
 \label{LBorel}[{\small Borel summation}]
 For any sequence $\{c_{i_1\ldots i_k}\}_{k\in\bf N}=\{c, c_i, c_{ij},\ldots\}$ there exists a smooth
function $f$ such that, for all $k\in \mathbf N$,
$$
  f- \sum_{p=0}^k c_{i_1\ldots i_p} y^{i_1}\cdots y^{i_p} = o_k(r^k)
   \;.
 $$
\end{lemma}

\noindent\proof
Let  $  \phi\in C^\infty({\bf R})$ be any function such that
$$
  \phi|_{[0,1/2]}=1\;, \quad  \phi|_{[1,\infty)}=0
 \;.
$$
Set $f_0=c$, and for $p>1$
\begin{equation}
 \label{19IX0.5}
 f_ p = \sum_{i_1,\ldots,i_p} \phi(M_p |y|) c_{i_1\ldots i_p} y^{i_1}\cdots y^{i_p}
  \;,
\end{equation}
where the constant $M_p$ is chosen large enough so that {for
all $p>0$ and for all multi-indices $\alpha$ satisfying}
$$
  0\le|\alpha|\le p-1 \ \mbox{we have} \  |\partial^\alpha f_ p | \le 2^{-p}
  \;.
$$
Then for each $\alpha$ the series
$$
  \sum_{p=0}^\infty \partial^\alpha f_p
$$
is absolutely convergent. By standard results (see, e.g.,
\cite[Volume~I, Theorem~1.1.5]{HormanderAPDE}), the function
$$
 f:=  \sum_{p=0}^\infty  f_p
$$
is smooth, and is easily seen to have the required properties.
\hfill $ \Box$

\bigskip

\noindent{\sc Acknowledgements:} Supported in part by the
Polish Ministry of Science and Higher Education grant Nr N N201
372736. PTC acknowledges many useful discussions with
Y.~Choquet-Bruhat and Jose-Maria Martin Garcia on problems
closely related to the ones addressed here.

%
%
%
%
%
%

\bibliographystyle{amsplain}
\bibliography{%
../references/reffile,%
../references/newbiblio,%
../references/newbiblio2,%
../references/bibl,%
../references/howard,%
../references/bartnik,%
../references/myGR,%
../references/newbib,%
../references/Energy,%
../references/netbiblio,%
../references/PDE}

\end{document}